\documentclass[aps,pra,twocolumn,superscriptaddress]{revtex4}

\usepackage{graphicx}
\usepackage{dcolumn}
\usepackage{bm}
\usepackage{times}
\usepackage{subfigure}
\usepackage{array}

\begin{document}


\title{Frequency metrology on the \begin{math}4s\,^2\mathrm{S}_{1/2}-4p\,^2\mathrm{P}_{1/2}\end{math}  transition in $^{40}$Ca$^+$ for a comparison with quasar data}


\author{A.L. Wolf}\email{alwolf@few.vu.nl}\affiliation{Laser Centre Vrije Universiteit,
De Boelelaan 1081, 1081 HV Amsterdam, The Netherlands}
\affiliation{NMi van Swinden Laboratorium, Thijsseweg 11, 2629 JA Delft, The Netherlands}
\author{S.A. van den Berg}
\affiliation{NMi van Swinden Laboratorium, Thijsseweg 11, 2629 JA Delft, The Netherlands}
\author{C. Gohle}\affiliation{Laser Centre Vrije Universiteit,
De Boelelaan 1081, 1081 HV Amsterdam, The Netherlands}
\author{E.J. Salumbides}\affiliation{Laser Centre Vrije Universiteit,
De Boelelaan 1081, 1081 HV Amsterdam, The Netherlands}
\author{W. Ubachs}\affiliation{Laser Centre Vrije Universiteit,
De Boelelaan 1081, 1081 HV Amsterdam, The Netherlands}
\author{K.S.E. Eikema}\affiliation{Laser Centre Vrije Universiteit,
De Boelelaan 1081, 1081 HV Amsterdam, The Netherlands}


\date{\today}

\begin{abstract}
High accuracy frequency metrology on the \begin{math}4s\,^2\mathrm{S}_{1/2}-4p\,^2\mathrm{P}_{1/2}\end{math} transition in calcium ions is performed using laser cooled and crystallized ions in a linear Paul trap. Calibration is performed with a frequency comb laser, resulting in a transition frequency of $f=755\, 222\, 766.2(1.7)$ MHz. The accuracy presents an improvement of more than one order of magnitude, and will facilitate a comparison with quasar data in a search for a possible change of the fine structure constant on a cosmological time scale. 
\end{abstract}

\pacs{32.30.-r, 42.62.Eh, 98.62.Ra, 37.10.Ty}

\maketitle


\section{Introduction\label{sect:intro}}
The detection of a possible variation in the fine structure constant
$\alpha$ by Webb {\it et al.}~\cite{webb} was founded on the so-called
many-multiplet method. In this study accurate wavelengths of absorption lines
of a number of atoms and ions, observed at high redshift in the
line-of-sight of Quasi Stellar Objects (QSO or `quasars'), were compared
with the laboratory (zero redshift) wavelengths of those atomic resonances.
Such an analysis may yield a value for $\Delta\alpha/\alpha$ over time
spans of many billion years. Improved methods and inclusion
of over a hundred of high-redshift absorption systems now has delivered
a positive effect with a $5\sigma$ significance \cite{murphy}.
The current opportunities for highly accurate wavelength calibrations of
astrophysical
data, provided by echelle-grating spectrometers attached to very large telescopes,
have brought about the need for renewed and updated laboratory precision
measurements. For several spectroscopic lines the absolute wavelength accuracy does not
match the level of $10^{-7}$, which is feasible in astrophysical
observations.
 One of the lines that need a more accurate frequency calibration is the $4s\,^2\mathrm{S}_{1/2}-4p\,^2\mathrm{P}_{1/2}$ transition in Ca$^+$, which is the subject of this paper \cite{berengut}.

In general, trapped calcium ions are studied both for use in very accurate ion clocks \cite{champenois,toyoda} and for quantum computation \cite{roos99,roos}. Nevertheless, the  $4s\,^2\mathrm{S}_{1/2}-4p\,^2\mathrm{P}_{1/2}$ transition that is used for Doppler cooling these ions, has been given far less attention. For this transition, often the value $\nu=25\,191.5182$  cm$^{-1}$  is cited \cite{morton}. This frequency is based on a private communication with U. {Litz{\'e}n}, who recently recalibrated the corresponding data yielding an updated value of $\nu=25\,191.519(0.002)$  cm$^{-1}$ \cite{litzen}. The next best wavelength calibration dates back as far as 1956, with a value $\nu=25\,191.51(0.13)$ cm$^{-1}$, obtained by {Edl\'en} and Risberg using classical spectroscopy \cite{edlen}.

In the present study, a linear Paul trap is used for confinement of a cloud of calcium ions, while laser cooling is employed to obtain a linewidth equal to twice the natural linewidth of the transition. The cold ions are probed using a second laser beam of lower intensity. Simultaneously, a part of this laser is heterodyned to a frequency comb for absolute frequency calibration. 

\section{Experimental setup and procedures \label{sect:setup}}

\subsection{Ion production and trapping \label{sect:ions}}
\begin{figure}
 \begin{center}
  \subfigure[\ Side view]{\label{fig:traptot}\includegraphics[width=40 mm]{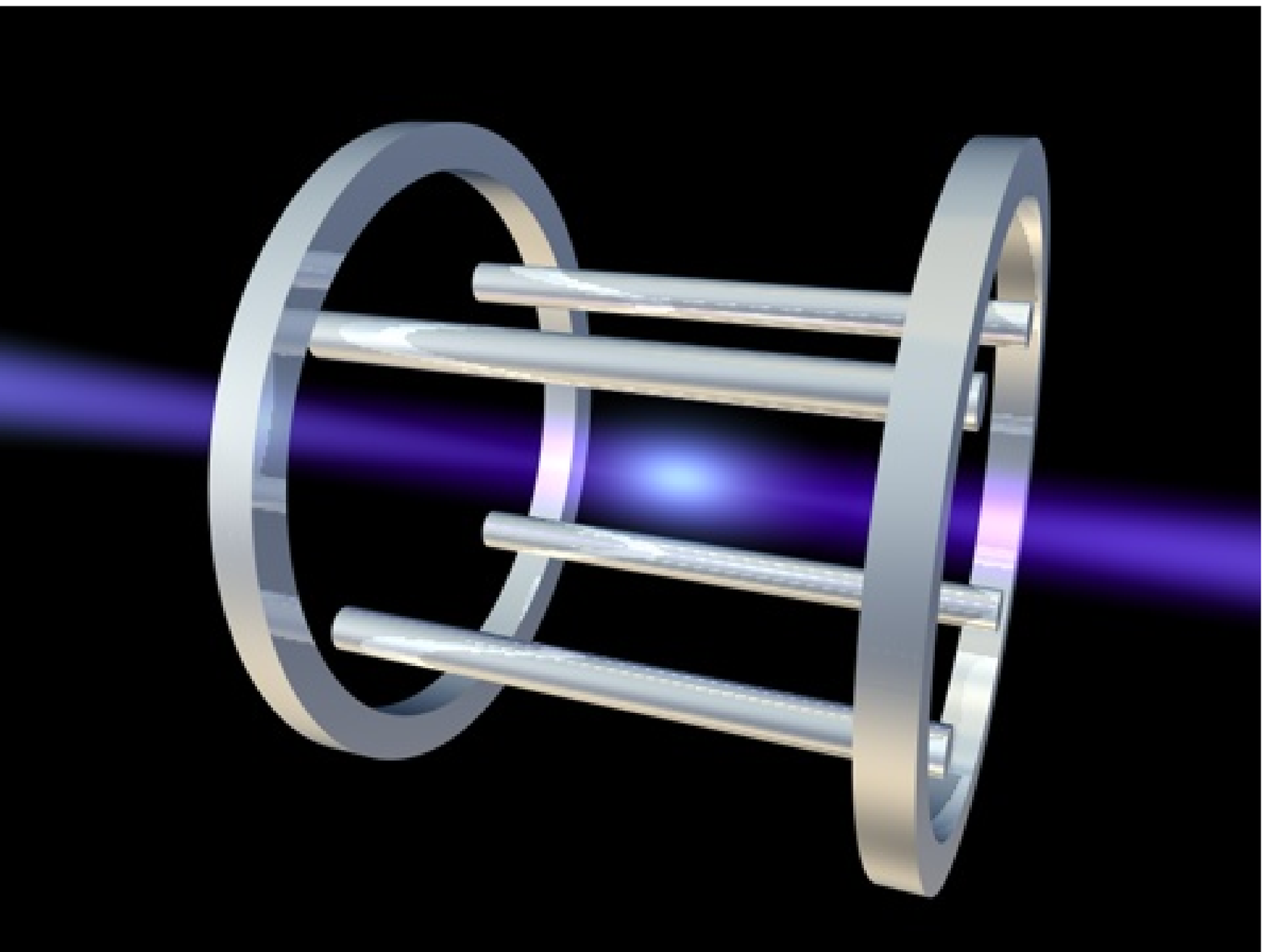}}%
  \subfigure[\ Axial view]{\label{fig:trapaxis}\includegraphics[width=50 mm]{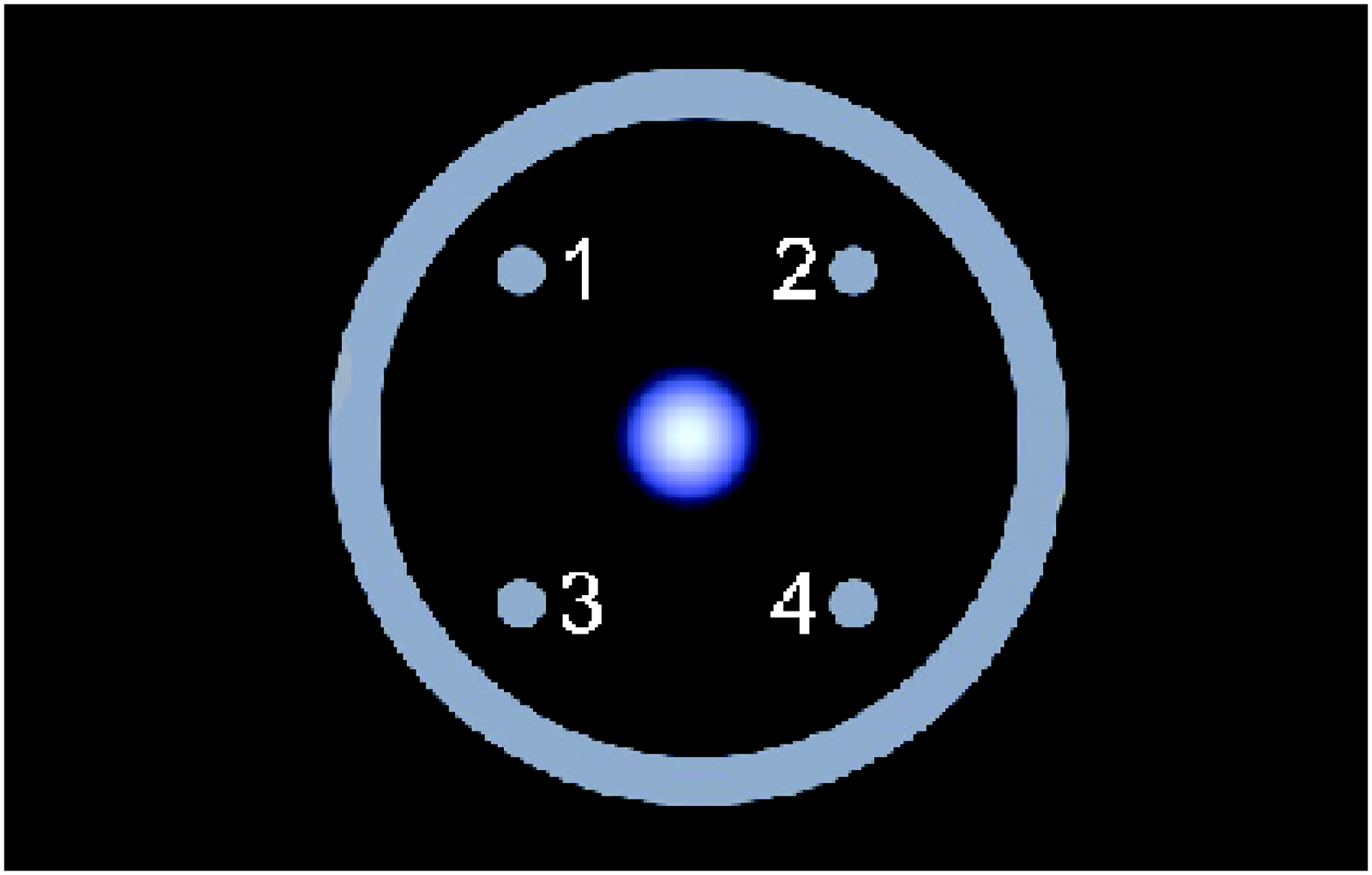}}
\end{center}
\caption{(Color online) Electrode configuration for the linear trap. The trap consists of four linear electrodes (labeled 1-4) for radial confinement: A Radio Frequency voltage is put on two diagonally facing rods (1,4), while the other two rods are grounded (2,3). A DC voltage is applied to the ring electrodes for axial confinement.\label{fig:trap}}
\end{figure}

The calcium ions are trapped in a linear Paul trap (Fig. \ref{fig:trap}) consisting of 4 linear electrodes for radial confinement, and 2 ring electrodes for restriction in the axial direction. These elements are suspended and separated by insulating spacers. The electrode dimensions enable fluorescence collection over a large solid angle of $\approx 0.2 \pi$ sr. The four linear electrodes are placed at a distance $r_0=4$ mm from the trap axis, and the ring electrodes are $17$ mm apart. A radio frequency (RF) voltage $V_0$ at a frequency $\Omega=2 \pi\times3.3$ MHz is applied to the electrodes 1 and 4, while the electrodes 2 and 3 are grounded.  The characteristic stability parameter of the Mathieu equation, which describes the motion of an ion in an harmonic RF potential, is $q=2QV_0/m\Omega^2r_0^2$ \cite{berkeland}. Here $Q$ and $m$ are the charge and mass of the ions respectively. A small stability parameter ($q<<1$) is advantageous to reduce RF heating \cite{prestage,ryjkov}. However, trap imperfections and micromotion in an extended cloud can still lead to substantial heating of the confined ions \cite{berkeland, zhang}. The trap is generally operated with a $V_0 \approx 60$ V, corresponding to $q=0.04$. The ring electrodes that are held at a constant voltage of typically \mbox{$V_{DC}=10$ V} create a harmonic potential for axial confinement.

%

\begin{figure}
\includegraphics[width=85mm]{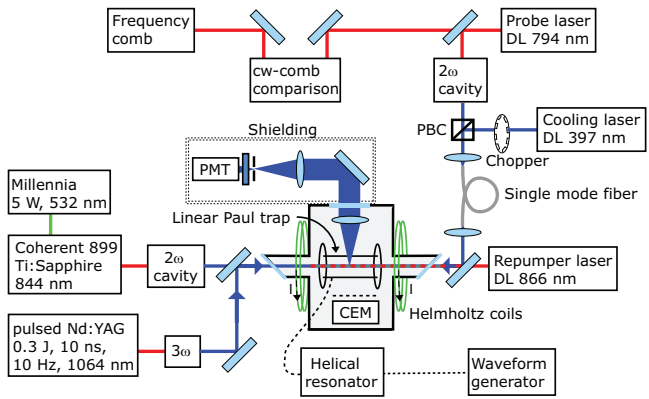}%
\caption{(Color online) Schematic design of the laser system and linear Paul trap.  DL=Diode laser, PMT=Photo-Multiplier Tube, 2$\omega$=frequency-doubling setup, 3$\omega$=frequency-tripling setup, PBC=Polarizing Beamsplitter Cube, CEM=Channel Electron Multiplier.\label{fig:setup}}
\end{figure}

The trap is installed in a vacuum chamber, which is evacuated to a pressure of $2\times10^{-9}$ mbar. Neutral calcium atoms are evaporated from an oven into the linear Paul trap. In the trapping region calcium atoms are resonantly excited to the 4$p\,^1$P$_1$ level by a frequency-doubled CW (continuous wave) Ti:sapphire ring laser (Coherent 899) producing 422 nm, and subsequently ionized by a frequency tripled Nd:YAG (yttrium aluminum garnet) laser operating at 355 nm (Fig. \ref{fig:setup}). The RF source for the trap (Agilent 33120A) is resonantly enhanced and filtered by a helical resonator (with a loaded Q$\approx$140). For diagnostic purposes a Channel Electron Multiplier (CEM) and accelerating grid were installed underneath the Paul trap. By pulsed extraction of the ions from the trap with a variable delay and detection on the CEM, the ionization rate and trapping conditions can be probed and optimized.

\subsection{Doppler cooling to ion cloud crystallization\label{sect:cool}}
\begin{figure}[!hbt]
\includegraphics[width=85mm]{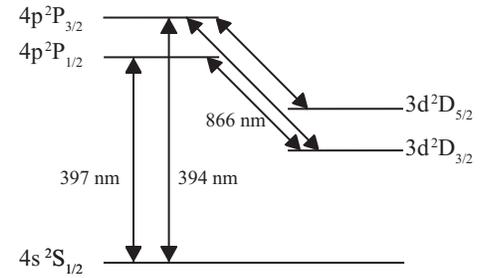}%
\caption{Partial energy level diagram of Ca$^+$\label{fig:calcium}}
\end{figure}

In order to counteract RF heating and reduce the Doppler width of the measured  transition, the ions are laser cooled on the \begin{math}4s\,^2\mathrm{S}_{1/2}-4p\,^2\mathrm{P}_{1/2}\end{math} transition (the relevant energy levels are shown in Fig. \ref{fig:calcium}). The cooling laser is a 397 nm grating stabilized diode laser ($P=3$ mW,  Toptica DL100).  Since the ions in the excited $4p\,^2\mathrm{P}_{1/2}$ state have a 7\% probability of decaying to the long-lived $3d\,^2\mathrm{D}_{3/2}$ state, an additional diode laser at 866 nm ($P=1$ mW, Toptica DL100) is used for repumping.

\begin{figure}[!hbt]
\includegraphics[width=60mm]{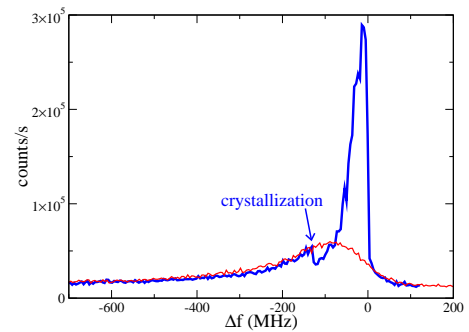}%
\caption{(Color online) Observed fluorescence vs. frequency detuning $\Delta f$ of the cooling laser relative to the $4s\,^2\mathrm{S}_{1/2}-4p\,^2\mathrm{P}_{1/2}$ transition in Ca$^+$. Red thin trace: $T_{\mathrm{ions}}\approx40$ K, blue thick trace: $T_{\mathrm{ions}}\approx4$ K. The sudden decrease in fluorescence in the blue curve is the typical signature for crystallization of the ion cloud.\label{fig:cooling}}
\end{figure}

Excitation of the transition is observed through the emitted fluorescence at 397 nm. Using a 25 mm diameter lens with a focal length of 25 mm inside the vacuum chamber, $5$ \% of this light is collected. The ion cloud is imaged onto a pinhole to remove background radiation and scattered light from the electrodes, and its fluorescence is detected on a photomultiplier tube (PMT, Philips XP2020Q). The cooling laser is spatially filtered by a single-mode optical fiber (reducing the effective cooling laser power to $P=1$ mW), before being focused into the trapping region. This further reduces the background signal due to scattered light from the trap electrodes, which mainly originates from the 397 nm cooling laser.  In Fig. \ref{fig:cooling} two excitation spectra of trapped and laser cooled ions are shown, obtained by scanning the cooling laser. Depending on the cooling efficiency (controlled by the repumper detuning), the recorded fluorescence follows either a red-shifted, distorted Gaussian profile (red curve), or shows a dip followed by a much narrower resonance (blue curve) which is typical for crystallization \cite{diedrich}. Such a crystallized ion cloud is used for the spectroscopy to minimize Doppler broadening. Laser cooling is performed with the cooling laser set at detuning $\Delta f \approx 10$ MHz from resonance (locked to a wavemeter (Atos model LM-007) within $\approx 6$ MHz), which is approximately half the natural linewidth. The achieved temperature is typically $T\approx 1$ K. This is higher than the Doppler limit of $T=0.5$ mK, probably due to anharmonicity of the applied electric field or other trap imperfections \cite{berkeland,zhang}. The large number of trapped ions ($\sim 100$) magnifies this effect, since the Coulomb repulsion forces the ions away from the more harmonic trap center. The lifetime of the ions in the trap is measured to be $\tau \approx 8$ min by monitoring the fluorescence from the cooling laser.

\subsection{Spectroscopy and absolute frequency calibration \label{sect:spectroscopy}}

Because the \begin{math}4s\,^2\mathrm{S}_{1/2}-4p\,^2\mathrm{P}_{1/2}\end{math} transition is both the cooling and measured transition, the cooling laser is periodically switched off using a chopper with a duty cycle of 0.5 at a frequency of 0.3 kHz. The fluorescence due to the spectroscopy laser is recorded during these periods. A diode laser at 794 nm (Toptica DL100) is frequency doubled to the 397 nm to probe the transition. To minimize heating of the cloud the probe laser intensity is a thousand times weaker than the cooling laser. After spatial filtering a  power of only $P\approx 1$ $\mu$W remains, which is  focused to 0.5 mm diameter in the trap. The induced fluorescence is detected with the imaging system and PMT described previously (Fig. \ref{fig:setup}). The fluorescence of both the cooling laser and spectroscopy are separately recorded, using a gating technique synchronized to the chopper in the cooling laser beam. The cooling and probe laser are guided through the same single mode fiber (Fig. \ref{fig:setup}), ensuring good alignment of the weak probe beam onto the cold ion cloud, while also spatially filtering the probe beam. In addition, by monitoring the cooling laser induced fluorescence, ion loss from the probed region can be observed during the scan.

The probe laser is calibrated on an absolute frequency scale by referencing it to a frequency comb \cite{holzwarth, diddams}. A frequency comb laser is characterized by two frequencies: The repetition frequency $f_{rep}$ which determines the mode spacing, and the carrier-envelope offset frequency $f_{ceo}$, that gives the comb its offset from zero frequency. Both frequencies are locked to a Rubidium atomic clock (Stanford Research Systems PRS-10), which is referenced to the Global Positioning System (GPS), fixing the frequency of the  $n^{th}$ mode  to $f_n = \pm f_{ceo}+n\times f_{rep}$. The Allan deviation of the frequency comb modes is determined from the stability of the repetition rate relative to the clock combined with the stability of the Rubidium frequency reference. For the typical averaging time of 1 second, the Allan deviation is $\sigma=0.1$ MHz for optical modes at $377$ THz, leading to an uncertainty of $0.2$ MHz in the calibration at \mbox{$755$ THz.} The comb laser operates at a repetition rate of 207 MHz and with a spectrum of 54 nm FWHM around 825 nm.

\begin{figure}[!hbt]
\includegraphics[width=60mm]{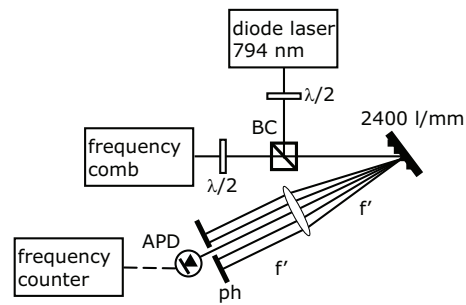}%
\caption{Schematic view of the setup for heterodyning the CW diode laser with the frequency comb. The CW laser is overlapped with the comb on a beamsplitter (BC). A grating-based spectral filter is used to reduce the number of comb modes on the detector. For clarity a lens is drawn, in practice a curved mirror is used. APD=Avalanche Photodiode, ph=pinhole, $\lambda/2$=half-wave plate. \label{fig:beat}}
\end{figure}

An interference beat note is generated between the spectroscopy laser and the comb laser modes, by overlapping the near-infrared fundamental output of the spectroscopy laser with the output of the frequency comb laser. The comb light is spectrally filtered using a grating and curved mirror in a 2f' configuration (Fig. \ref{fig:beat}) to reduce the amount of comb modes and improve the signal to noise ratio of the recorded beat note with the spectroscopy laser. An  avalanche photodiode (Menlo Systems APD210) detects the 40 dB (bandwidth 400 kHz) beat note $f_{bn}$ between the two lasers, which is monitored using a frequency counter (Agilent 53132A). 

The laser frequency $f_l$ follows from the locked frequencies $f_{ceo}$, $f_{rep}$, and the measured beat frequency $f_{bn}$ through the equation:

\begin{equation}
f_l = \pm f_{bn}+ \pm f_{ceo}+n\times f_{rep}
\end{equation}

The sign of $f_{bn}$ can be simply deduced from the known scanning direction of the laser. The scan over the transition was repeated at a different locking frequency of the carrier envelope offset frequency in order to determine the sign of $f_{ceo}$. In order to calibrate the laser frequency the mode number $n$ still has to be determined. The most accurate measurement up to date for the $4s\,^2\mathrm{S}_{1/2}-4p\,^2\mathrm{P}_{1/2}$ transition in Ca$^+$ has an uncertainty of $60$ MHz, which complicates assigning the mode number $n$ with confidence. This is solved by repeating the measurement at two different comb repetition frequencies, sufficiently far apart to distinguish between the adjacent mode numbers. As the transition frequency is independent of the used comb repetition rate, the measurements coincide for a particular mode number $n$, from which the absolute frequency can be determined (see below).

\begin{figure}[!hbt]
\includegraphics[width=85mm]{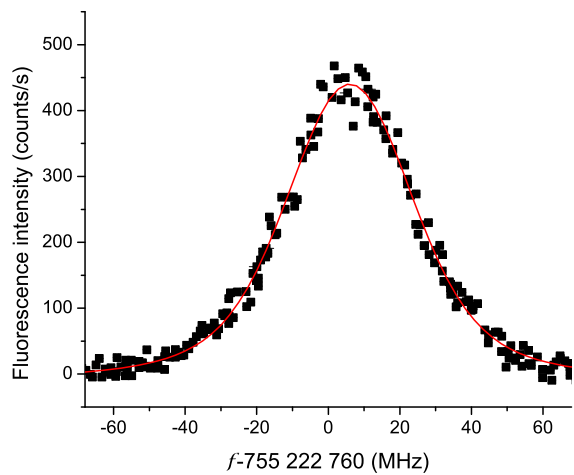}%
\caption{(Color online) An example of the measured fluorescence spectrum for the $4s\,^2\mathrm{S}_{1/2}-4p\,^2\mathrm{P}_{1/2}$ transition in Ca$^+$ (black dots) and the corresponding Voigt fit (red line), corrected for the trap loss during the scan. \label{fig:scan}}
\end{figure}

\subsection{Results}

A typical scan of the spectroscopy laser takes about 2 minutes, which is substantial compared to the trap lifetime of $\approx8$ min. In order to minimize distortion of the line profile, the spectroscopy signal is corrected for this ion loss. Since the cooling is insufficient to reach the natural linewidth of the transition, a Voigt profile, i.e. a convolution of a Gaussian and a Lorentzian is fitted to the data. The width of the Lorentzian is fixed to 22.4 MHz \cite{jin}, the natural linewidth of the transition. The width of the Gaussian component is fitted, and typically on the order of $20$ MHz. From this fitted value an ion cloud temperature of $T\approx1$ K is deduced (Fig. \ref{fig:scan}).

\begin{figure}[!hbt]
\includegraphics[width=80mm]{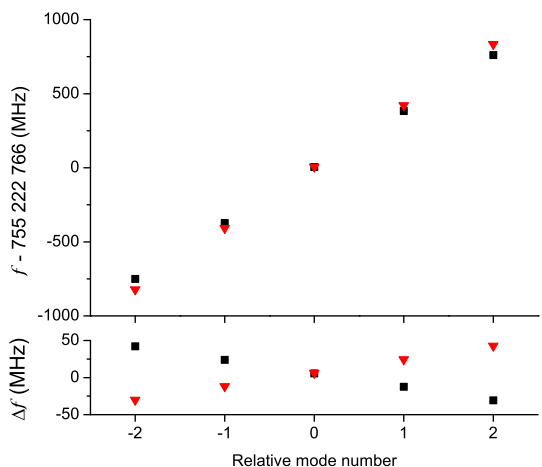}%
\caption{(Color online) Absolute calibration of the   $4s\,^2\mathrm{S}_{1/2}-4p\,^2\mathrm{P}_{1/2}$ transition in Ca$^+$ is performed by finding the coincidence of measurements with two different frequency comb repetion rates of 189 MHz (circles) and 207 MHz (triangles). The depicted transition frequency $f=2\times f_{l}$, since the laser frequency is measured in the infrared, after which the laser is frequency doubled for the spectroscopy. Consequently the spacing between adjacent points in the picture is $2\times f_{rep}$. The upper graph shows the frequency dependence on the choice of mode number. The lower graph shows the frequency difference between the points in the upper graph, with $\Delta f=f-755\,222\,766$ MHz$-2\times n' \times 198$ MHz, where $n'$ is the relative mode number (error bars are smaller than the point size). \label{fig:mode}}
\end{figure}

The mode number is determined by measuring the transition frequency using different frequency comb repetition rate of $f_{rep}=189$ MHz and $f_{rep}=207$ MHz. This results in two sets of possible transition frequencies (Fig. \ref{fig:mode}). The measurements clearly coincide at a value of $f=755\,222\,766$ MHz, which is near the literature value of $f=755\,222\,740(60)$ MHz.

Several effects that can introduce systematic and statistical errors on the transition frequency have been investigated. These include the Stark shifts caused by both the light field and the Paul trap, the stability of the comb modes and the Zeeman shift caused by the earth magnetic field. The respective shifts and statistical uncertainties are tabulated in \mbox{Table \ref{table:errors}.}

\begin{figure}[!hbt]
\includegraphics[width=85mm]{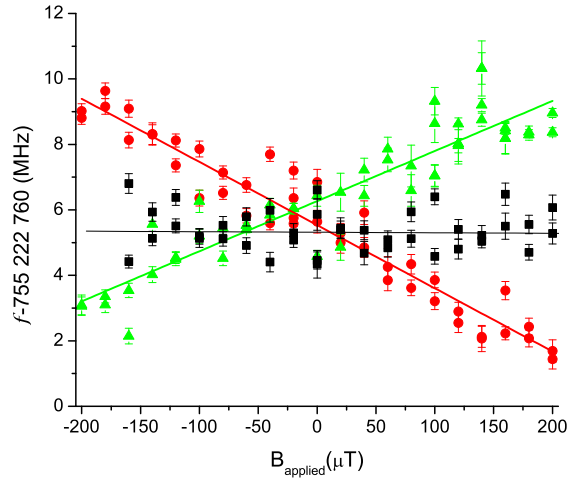}%
\caption{(Color online) Transition frequency for the $4s\,^2\mathrm{S}_{1/2}-4p\,^2\mathrm{P}_{1/2}$ transition in Ca$^+$ with respect to an applied magnetic field $B_{applied}$. Three different polarizations were used (see text): right-circular ($\sigma^+$, green triangles), left-circular ($\sigma^-$, red circles) and linearly polarized light ($\sigma^+ + \sigma^-$, black squares).\label{fig:B_ext}}
\end{figure}

In order to probe the Zeeman shift, a magnetic field is generated in the direction of the laser beam using Helmholtz coils (9 windings, diameter $2r=19$ cm). For magnetic fields in this direction, a deviation from linear spectroscopy laser polarization can lead to an unequal excitation of the $\Delta m = \pm  1$ transitions. The Zeeman shift in the involved energy levels ($\Delta E=g_Jm_J\mu_B B$) results in a magnetic field dependent shift in the transition frequency of $\pm 19$kHz/$\mu$T for the $\Delta m= \pm 1$ transitions. For perpendicular magnetic field components no shifts of the line center, but only broadening can be expected to first order. These are therefore not considered.

The line center of the transition as a function of the applied external fields was recorded for three different polarizations: right- and left-circular ($\sigma^+$ and $\sigma^-$ respectively) and linear ($=\sigma^+ + \sigma^-$). The result is shown in Fig. \ref{fig:B_ext}. Linear fits to the data show Zeeman coefficients of $19(1)$kHz/$\mu$T and $15(1)$kHz/$\mu$T for the $\sigma^+$ and $\sigma^-$ components. The deviation of the latter from the calculated value can be attributed to an inaccurate control of the polarization in the interaction region, due to depolarization by the Brewster angled vacuum windows. For excitation with linear polarization, a first order Zeeman coefficient of $0.0$ kHz/$\mu$T is expected, and it was correspondingly measured to be $0.2(1.0)$ kHz/$\mu$T. Therefore the final calibration of the $4s\,^2\mathrm{S}_{1/2}-4p\,^2\mathrm{P}_{1/2}$ transition is based on an average over the measurements with linear polarization. In this case, for all applied magnetic fields, the Zeeman shift is smaller than $200 \times 0.2(1.0)$ kHz $=0.04(0.2)$ MHz. The crossing of the three curves yields the zero total magnetic field at $B_{applied}=21(6) \mu$T (in good agreement with the expected earth magnetic field component of $19 \mu$T). From this value an upper limit for the total residual Zeeman shift is calculated of $0.04(3)$ MHz, which is negligible.

\begin{table}
\caption{Measured systematic shifts and standard uncertainty budget ($1\sigma$). All values in MHz.\label{table:errors}}
\begin{ruledtabular}
\begin{tabular}{|l|r|r|}\hline
Effect & Shift (MHz) & $1\sigma$ Uncertainty (MHz)  \\ \hline
Zeeman  &  0.0 & 0.0\\
AC Stark repumper & -0.4 & 0.6\\
AC Stark spectroscopy laser & -0.4 & 0.8\\
RF Stark effect &  0.0 & 1.2 \\ 
Comb calibration & 0.0 & 0.2 \\
Statistics &   & 0.6\\\hline
Total  &  -0.8 & 1.7\\

\end{tabular}
\end{ruledtabular}
\end{table}

AC Stark shifts in the spectroscopy data for both the repumper laser and the probe laser were evaluated by changing their respective intensities. As expected, both shifts were found to be small compared to the measurement accuracy ($-0.4(0.6)$ MHz and $-0.4(0.8)$ MHz respectively, Table \ref{table:errors}). In order to assess possible trap induced shifts, the RF voltage $V_0$ was varied over a range from 50-200 $V$. The corresponding shift was measured to be consistent with a zero value: $\Delta f = 0.0 (1.2)$ MHz.

The $^{40}$Ca$^+$ $4s\,^2\mathrm{S}_{1/2}-4p\,^2\mathrm{P}_{1/2}$ transition frequency follows from the statistical average of the measurements with linearly polarized light, corrected for the measured shifts. The total measured value adds up to $f=755\,222\,765.4(0.6)+0.4(0.6)+0.4(0.8)+0.0(1.2)+0.0(0.2)=755\,222\,766.2(1.7)$ MHz, with the uncertainties taken in quadrature. The present result is consistent with the most accurate previously reported value of $\nu=25\,191.519(0.002)$ cm$^{-1}$ ($f=755\,222\,740(60)$ MHz).



\section{Conclusions\label{sect:conclusion}}

We have performed an absolute frequency calibration of the $4s\,^2\mathrm{S}_{1/2}-4p\,^2\mathrm{P}_{1/2}$ transition in $^{40}$Ca$^+$ ions in a linear Paul trap, including a study of Zeeman and Stark shifts. The transition was measured to be at \hbox{$f=755\,222\,766.2 (1.7)$ MHz}, in a laser cooled ion cloud of $T\approx 1$ K. The presented value represents more than an order of magnitude improvement over the previously most accurate measurement. This level of accuracy, at $\Delta \lambda / \lambda=2\times  10^{-9}$, is such that for any comparison with state-of-the-art astrophysical data, the laboratory value can be considered exact.

The employed technique for trapping and cooling ions will be used for spectroscopy on other ions in the near future. We plan to trap and sympathetically cool ions such as Mn$^+$ and Ti$^+$ with Ca$^+$. In this way, we expect to reach improvements in energy level calibration of these ions, similar to the demonstrated accuracy for calcium. This will provide accurate laboratory wavelengths for the comparison with high redshift astronomical data in a search for a possible variation of the fine structure constant over cosmological time scales.


\begin{acknowledgments}
This project is financially supported by the Foundation for
Fundamental Research on Matter (FOM), and the Nederlands Meetinstituut (NMi). These contributions are gratefully
acknowledged.
\end{acknowledgments}

%




\begin{thebibliography}{99}


\bibitem{webb} J.K. Webb, V.V. Flambaum, C.W. Churchill, M.J. Drinkwater, J.D. Barrow,
Phys. Rev. Lett. \textbf{82}, 884-887 (1999).

\bibitem{murphy} M. T. Murphy, J. K. Webb, V. V. Flambaum,
 Mon. Not. R. Astron. Soc. \textbf{345}, 609-638 (2003).




\bibitem{berengut} J.C. Berengut, V.A. Dzuba, V.V. Flambaum, M.V. Marchenko, J.K. Webb,
ArXiv: physics.atom-ph \textbf{}, 0408017 (2006).

\bibitem{champenois} C. Champenois, M. Houssin, C. Lisowski, M. Knoop, G. Hagel, M. Vedel, F. Vedel,
Phys. Lett. A \textbf{331}, 298-311 (2004).

\bibitem{toyoda} K. Toyoda, H. Naka, H. Kitamura, H. Sawamura, S. Urabe, 
Opt. Lett. \textbf{29}, 1270-1272 (2004).

\bibitem{roos} C.F. Roos, M. Riebe, H. Haffner, W. Hansel, J. Benhelm, G.P.T. Lancaster, C.Becher, F. Schmidt-Kaler, R. Blatt,
Science \textbf{304}, 1477-1480 (2004).

\bibitem{roos99} C.F. Roos, T. Zeiger, H. Rohde, H.C. Nagerl, J. Eschner, D. Leibfried, F. Schmidt-Kaler, R. Blatt,
Phys. Rev. Lett. \textbf{83}, 4713-4716 (1999).


\bibitem{morton} D.C. Morton,
Astrophys. J. Suppl. Ser. \textbf{151}, 403 (2004).

\bibitem{litzen} U. {Litz\'en}, Private communication (2008)


\bibitem{edlen} B. {Edl\'en}, and P. Risberg,
Ark. Fys. \textbf{10}, 553-566 (1956).

\bibitem{berkeland} D.J. Berkeland, J.D. Miller, J.C. Bergquist, W.M. Itano, D.J. Wineland,
J. Appl. Phys. \textbf{83}, 5025-5033 (1998).

\bibitem{prestage} J.D. Prestage, A. Williams, L. Maleki, M.J. Djomehri, E. Harabetian,
Phys. Rev. Lett. \textbf{66}, 2964-2967 (1991).

\bibitem{ryjkov} V.L. Ryjkov, X.Z. Zhao, H.A. Schuessler,
Phys. Rev. A \textbf{71}, 033414 (2005).

\bibitem{zhang} C.B. Zhang, D. Offenberg, B. Roth, M.A. Wilson, S. Schiller,
Phys. Rev. A \textbf{76}, 012719 (2007).

\bibitem{diedrich} F. Diedrich, E. Peik, J.M. Chen, W. Quint, H. Walther,
Phys. Rev. Lett. \textbf{59}, 2931-3934 (1987).

\bibitem{diddams} S.A. Diddams, D.J. Jones, J. Ye, S.T. Cundiff, J.L. Hall, J.K. Ranka, R.S. Windeler, R. Holzwarth, T. Udem, T.W. {H\"ansch},
Phys. Rev. Lett. \textbf{84}, 5102-5105 (2000).
 
\bibitem{holzwarth} R. Holzwarth, T. Udem, T.W. {H\"ansch}, J.C. Knight, W.J. Wadsworth, P.S.J. Russell,
Phys. Rev. Lett. \textbf{85}, 2264-2267 (2000).

\bibitem{jin} J. Jin and D.A. Church,
Phys. Rev. A \textbf{49}, 3463-3471 (1994).



\end{thebibliography}

\end{document}